\documentclass[prd,superscriptaddress,10pt,twocolumn,preprintnumbers]{revtex4}
\usepackage{color,graphicx,ulem}
\usepackage{amsmath,amssymb}
\usepackage{esint}
\begin{document}

\title{A note on the semiclassicality of cosmological perturbations}

\author{Pietro Don\`a}
\email[]{pietro_dona@fudan.edu.cn}
\affiliation{Department of Physics \& Center for Field Theory and Particle Physics, Fudan University, 200433 Shanghai, China}
\author{Antonino Marcian\`o}
\email[]{marciano@fudan.edu.cn}
\affiliation{Department of Physics \& Center for Field Theory and Particle Physics, Fudan University, 200433 Shanghai, China}

\date{\today}

\begin{abstract}
\noindent 
Moving from the consideration that matter fields must be treated in terms of their fundamental quantum counterparts, we show straightforward arguments, within the framework of ordinary quantum mechanics and quantum field theory, in order to convince readers that cosmological perturbations must be addressed in term of the semiclassical limit of the expectation value of quantum fields. We first take into account cosmological perturbations originated by a quantum scalar field, and then extend our treatment in order to account for the expectation values of bilinears of Dirac fermion fields. The latter can indeed transform as scalar quantities under diffeomorphisms, as well as all the other bilinear of the Dirac fields that belong to the Clifford algebra. This is the first of a series of works that is intended to prove that cosmological quantum perturbations can actually be accounted for in terms of Dirac fermion fields, which must be treated as fundamental quantum objects, and their dynamics. 
\end{abstract}

\maketitle

\section{Introduction}

\noindent
The theory of cosmological perturbations has been so far successfully addressed in term of quantum fluctuations of scalar fields over their classical background \cite{Kodama:1985bj, Mukhanov:1990me}. Despite the huge success of this treatment, a big question still remains usually unaddressed within the literature of cosmological perturbations: how can our current understanding of matter in the standard model of particle physics, and in particular the way in quantum field theory we deal with fermionic matter that are fundamental quantum, can be 
reconciled with this semiclassical framework? How can we account for a semiclassical background of quantum fields and consider perturbations as their quantum fluctuations?
We emphasize that this is not a mere academic question, since a naive answer about this issue will clearly imply that linear perturbations induced by spinorial fields are vanishing. And while the latter would be one of the most tangible consequences, we can be confident that it will not be the only one. Nonetheless, in order to show that linear perturbations can be originated also by bilinear fermionic operator, we must first clarify the quantum content of the perturbations of scalar matter fields. We intend to show then that what originates the cosmological perturbations, as addressed so far within the literature, are matter fields dealt with in the framework of quantum field theory. But differently from the previous literature, perturbations will be originated by the expectation value of quantum matter field operators on macroscopic condensate states that carry a perturbed spectrum of particles. Background quantities will correspond instead to the expectation value of the matter operators on macroscopic states with unperturbed spectrum of particles. We will then extend the arguments we will have recovered for scalar matter fields, to cosmological perturbations induced by fermionic matter, and show that the latter are actually non-vanishing at the linear order too. 

We are aware of the novelty of this analysis, and that conceptual difficulties might originate while shifting away from the usual semiclassical considerations deployed within the standard theory of cosmological perturbations. Nevertheless we are confident that our readers, only relying on their knowledge of quantum mechanic and standard quantum field theory, will appreciate the intrinsic consequences of this approach, which we emphasize again are not merely academic, but rather phenomenologically important. We emphasize that within this approach we are able to address questions raised so far on the quantum-to-classical transition of cosmological perturbations \cite{Kiefer:2008ku}, and that this is possible thanks to the new perspective we developed, which allows to overcome issues and shortcomings already summarized in the literature \cite{Sudarsky:2009za, Pinto-Neto:2013npa} and tackle fundamental questions on the quantumness of primordial cosmological perturbations and its detectability \cite{Martin:2015qta}.

At the same time, we acknowledge that many concepts we developed in this analysis, which are actually mutated from condensed matter theory, first appeared in the literature of cosmology in seminal papers by Brandenberger and Zhitnitsky \cite{Brandenberger:1996mr}, and by Alexander and Calcagni \cite{Alexander:2008yg}  (see also \cite{Alexander:2008vt, Alexander:2009uu}). Especially in the line of thought Alexander and Calcagni sought, the instantiation of the fundamental work of Bardeen, Cooper and Schrieffer on superconductivity \cite{BCS} enabled to find BCS like condensate even within the cosmological framework. These results have been achieved deploying a gravitational version of the Nambu--Jona-Lasinio mechanism, and the condensate hence obtained has been shown to play a crucial role for early or late cosmology. 

Following this path, but developing these ideas more on the Hamiltonian approach side than in the covariant path-integral formulation, we will show that it is possible to find macroscopic semiclassical states for both the bosonic and fermionic matter sectors. What we then call condensates, following the jargon of condensed matter, are actually states of the Hilbert space of the theory, respectively bosonic and fermionic, which are coherent in that they minimize the uncertainty relations between conjugated variables. In this study macroscopic states were merely addressed at the kinematical level. Nonetheless, in a forthcoming work \cite{BhramaDonaMarciano} we will show which effects arise from considering the dynamics, focusing on the phenomenological consequences and restrictions induced by the latter. 

We notice that such an interpretation is implicit is several investigations recently deepened in the literature of inflation \cite{Alexander:2011hz, Alexander:2014uza} and dark energy \cite{Dona:2015xia, Addazi:2016sot,Alexander:2016xbm, Addazi:2016nok}. The unquestionable novelty of this analysis stands anyway in linking the semiclassical limit of the quantum theory, and the macroscopic state of matter, to the development of a new setting for addressing the cosmological perturbations. The latter are then the by-product of the assumption of semi-classicality, and arise from the perturbations of the distributions in the momentum space that enter the macroscopic states.

The plan of the paper is the following. In Sec.~II we introduce macroscopic states of matter for bosonic matter fields, we specify their generalization and discuss their physical meaning. In Sec.~III we switch to the discussion of cosmological perturbations in the bosonic sector: we introduce a general framework to derive cosmological perturbations from the perturbation of the number density in the macroscopic coherent states of matter; we construct a quantum operator whose expectation value in the coherent perturbed states corresponds to the curvature perturbation variable; we finally outline how to derive standard results. In Sec.~IV we introduce macroscopic coherent states for fermionic matter, and specify the difference of our procedure with respect to bosonization. We then focus on the well known BCS states, and their SU(2) coherent states equivalents. In Sec.~V we develop, on the same foot of Sec.~III, a theory of cosmological perturbations that account for linear contributions from the fermionic sector. In Sec.~VI we show how number densities of macroscopic states transform under diffeomorphisms, and prove that coherent states are mapped into coherent states. In Sec.~ VII we spell conclusions and remarks. Detailed appendices follow on coherent states in the bosonic and fermionic sectors, on the relation between Bogolubov transformations and the adjoint action of the displacement operators, on the cosmological perturbations, and on the phenomenological observable which are sensitive to our analysis. 

\section{Macroscopic states of matter: scalar fields}

\noindent
Quantum mechanics (QM) is the fundamental framework we rely on to understand Nature \cite{Landau:1991wop, Sakurai:2011zz, CohenTannoudji:1987bi}. No disproval of this very fundamental framework have been recovered so far, and experimental data do actually confirm us in our every day life that quantum mechanics must not be questioned yet. 
We then start taking into account the states whose fluctuations of the number operator are negligible for a large number of quanta within the system that is considered. These are the coherent states \cite{CohenTannoudji:1987bi}, and represent a macroscopic wave-function that takes a special role in recovering the semiclassical limit \cite{Mandl:1985bg} of quantum mechanical operators in quantum field theory (QFT). 

\subsection{Coherent state for scalar fields}
\noindent
For the purpose of simplicity in what follows we will treat the case of a free real scalar field on flat (Minkowski) background, whose density Lagrangian and Hamiltonian in natural units read respectively 
\begin{eqnarray}
L(x)&=& \partial_\mu \phi(x) \partial^\mu \phi(x)\,, \\
H(x)&=& \pi^2(x)+\nabla \phi(x) \cdot \nabla \phi(x) \,, \nonumber
\end{eqnarray}
having introduced the conjugated momentum $\pi(x)=\dot{\phi}(x)$ to $\phi(x)$, in which the dot denotes derivative with respect to time. The field $\phi(x)$ (and similarly its conjugated momentum) is decomposed in the Fock basis of the harmonic linear oscillators, as a superposition of creation and annihilation operators for each mode: 
\begin{eqnarray} \label{scafi}
\phi\left(x\right)=\int_k
\left(a_{k}e^{-ikx}+a_{k}^{\dagger}e^{+ikx}\right)\,,
\end{eqnarray}
where the integration over the momentum space has to be understood with the appropriate measure. We then naturally extend the definitions of the quantum mechanical harmonic oscillator coherent state (see the Appendix I for more details), and consider the bosonic coherent state, labelled by the function $\alpha(k)\,:\,\mathbb{R}^{3}\to\mathbb{C}$, 
\begin{eqnarray}
\left|\alpha\right\rangle &\equiv&\prod_{k}\left|\alpha\left(k\right)\right\rangle =\prod_{k}e^{\alpha\left(k\right)a_{k}^{\dagger}-\alpha^{*}\left(k\right)a_{k}}\left|0\right\rangle
\\
&=&e^{\int d^{3}k\,\alpha\left(k\right)a_{k}^{\dagger}-\alpha^{*}\left(k\right)a_{k}}\left|0\right\rangle= D\left(\alpha\right)\left|0\right\rangle\,. \nonumber
\end{eqnarray}
The displacement operator $D\left(\alpha\right)=\exp({\alpha a^{\dagger}-\alpha^{*}a})$ inherits all the property of the harmonic oscillator counterparts (see {\it e.g.} Appendix I), in particular it is unitary and its action on a creation operator is 
\begin{align} \label{btri}
&D\left(\alpha\right)^\dagger a_k D\left(\alpha\right) = a_k + \alpha(k)\, , \\
& D\left(\alpha\right)^\dagger a^\dagger_k D\left(\alpha\right) = a^\dagger_k + \alpha^*(k)\, .
\end{align}
One trivially obtains that the classical real scalar field in terms of the function $\alpha$ is expressed as
\begin{eqnarray}
\phi_{\alpha}\left(x\right)&\equiv&\left\langle \alpha\right|\phi\left(x\right)\left|\alpha\right\rangle\nonumber \\
&=&\int_k
\left(\alpha_{k}e^{-ikx}+\alpha^{*}{}_{k}e^{+ikx}\right)\,.
\end{eqnarray}
Then the action of the displacement operator on the scalar field itself can be expressed in terms of the ``classical'' field $\phi_\alpha$.
\begin{eqnarray}
D\left(\alpha\right)^\dagger \phi \left(x\right) D\left(\alpha\right) = \phi \left(x\right)+ \phi_{\alpha}\left(x\right)\,.
\end{eqnarray}
It is useful to relate the expectation values of operators on a coherent state with vacuum expectation values of a the transformed operator
\begin{align}
&\left\langle \alpha\right|\phi\left(x_1\right)\ldots \phi\left(x_n\right) \left|\alpha\right\rangle \\
&\  = \left\langle 0\right| D^\dagger\left(\alpha\right) \phi\left(x_1\right) 
\ldots \phi\left(x_n\right) D\left(\alpha\right) \left|0\right\rangle \\
&\ = \left\langle 0\right| \left(\phi\left(x_1\right) + \phi_\alpha \left(x_1\right) \right) 
\ldots \left(\phi\left(x_n\right) + \phi_\alpha \left(x_n\right) \right) \left|0\right\rangle\, ,
\end{align}
or more in general 
\begin{align}
\label{EVoperator}
&\left\langle \alpha\right| \mathcal{O} \left(\phi\left(x\right)\right) \left|\alpha\right\rangle  = \left\langle 0\right| \mathcal{O} \left(\phi\left(x\right)+\phi_\alpha\left(x\right)\right) \left|0\right\rangle\, .
\end{align}
Furthermore the expectation value of a normal ordered operator on a coherent state is exactly its classical value
\begin{align}
\label{EVoperatorNO}
&\left\langle \alpha\right| : \mathcal{O} \left(\phi\left(x\right)\right) : \left|\alpha\right\rangle  =  \mathcal{O} \left(\phi_\alpha\left(x\right)\right)\,.
\end{align}
\noindent The energy density of the system on such a state immediately follows, once the dispersion relation $E_{k}=\sqrt{\vec k^2}$ is recovered from the classical equations of motion, namely
\begin{eqnarray}
\frac{1}{V}\int_V \left\langle \alpha\right|H\left(x\right)\left|\alpha\right\rangle =\int\frac{d^{3}k}{\left(2\pi\right)^{3}}E_{k}\left|\alpha_k\right|^{2}\,,
\end{eqnarray}
in which the integral $\int_V$ is over a fiducial volume $V$ that is finally send to infinite. 

\subsection{Generalized coherent state and scalar fields} \label{sh}
\noindent
The coherent state construction we illustrated in the previous section can be readily generalized to the simplest compact group $SU(2)$ by utilizing the Schwinger representation of its Lie algebra. Let's consider first the Hilbert space of two harmonic oscillators spanned by the creation (annihilation) operators $a^\dagger_1$, $a^\dagger_2$ ($a_1$, $a_2$). On this Hilbert space we can define the following operators
\begin{equation}
J^a\equiv \left(\tau^a\right)^{\alpha\beta} a^\dagger_\alpha a_\beta\,,
\end{equation}
where $\tau^a$ are the $SU(2)$ generators, $a=1,2,3$ and $\alpha,\beta=1,2$. It is straightforward to verify that 
\begin{equation}
\left[J^a,J^b\right]= \left[\tau^{a},\tau^{b}\right]^{\alpha\beta} a_{\alpha}^{\dagger}a_{\beta}= i \epsilon^{abc} J_c
\end{equation}
generates a $SU(2)$ algebra. Following the construction described in detail in Appendix II, it is immediate to construct a $SU(2)$ coherent states. Since a scalar field contains infinitely many harmonic oscillators, it is sufficient to choose how to couple the oscillators ({\it e.g.} we can fix a momentum $\vec{p}$, then for each momentum $\vec{k}$ we can pick the couple $a_{\vec{k}}$ and $a_{\vec{k}+\vec{p}}$). For each couple of modes we can finally define a $SU(2)$ coherent state, and consider the tensor product of all of them for our purposes.

Furthermore, the same very prescription is generalizable to any $SU(N)$ \cite{Perelomov,SUn}

\subsection{Off-diagonal long range order and zero mode}
\noindent 
Let us now focus on the (Hadamard) one-particle density matrix evaluated on the coherent state $|\alpha\rangle$, which is expressed as the Fourier transform of the momentum distribution $N_k=\langle a_k^\dagger a_k \rangle$ by 
\begin{eqnarray}
\rho_{\rm 1-p} (x - x')&=& \int_{k,k'} 
e^{-\imath ( k  x- k'  x') } \langle a_{k}^\dagger a_{k'} \rangle\nonumber\\
&=&\int_k 
e^{-\imath E_k (t-t')} e^{\imath \vec k \cdot (\vec x - \vec x')} \langle a_k^\dagger a_k \rangle\,. \nonumber
\end{eqnarray}
We are dealing with a coherent state that is picked around a certain macroscopic value $k_0$, whose occupation number is a macroscopic number $N_0=|\alpha_{k_0}|^2$ such that all the other $|\alpha_{k}|$ are small. This coherent state will have a momentum distribution 
\begin{eqnarray} \label{N}
N_k=N_0 \delta(k,k_0) + n(k)\,,
\end{eqnarray}
in which with $n(k)$ we denote a smooth function of $k$. The density matrix now reads 
\begin{eqnarray}
\rho_{\rm 1-p} (t-t';\vec x - \vec x')=\frac{N_0}{V} + \int_k 
e^{-\imath \vec k \cdot (\vec x - \vec x')} \,n(k)\,. \nonumber
\end{eqnarray}
The constant contributions to $\rho_{\rm 1-p} (t-t';\vec x - \vec x')$ represents a condensate, labelled by $n_0\equiv N_0/V$. There exist coherent states endowed with a sufficiently smooth $n(k)$ such that in the limit of large $||x-x'||$ (here the norm must be intended  as the distance in a Minkowski flat space-time)
\begin{eqnarray}
\lim_{||x-x'|| \rightarrow \infty}  \rho_{\rm 1-p} (t-t';\vec x - \vec x') = \langle \phi(x) \phi(x')\rangle_0 \equiv n_0\,. \nonumber
\end{eqnarray}
This is a natural extension of the concept of off-diagonal long ranged
order (ODLRO) \cite{OliverPenrose, CNY}. For superfluid the interpretation is rather straightforward, because of the quantum coherence of the condensate, and has to do with the quantum mechanical amplitude of a process in which a particle is annihilated at $\vec x$, where it gets absorbed into the condensate, and another one is created at $\vec x'$, where it exits the condensate. Nonetheless, exactly as for a superfluid one expects that at large space distances quantum correlations must be suppressed, we expect for the relativistic system under scrutiny that in the limit $||x-x'|| \rightarrow \infty$ the expectation value of the product of fields as space-time points far a part behave like the expectation value of the product of the fields:
\begin{eqnarray}
\lim_{||x-x'|| \rightarrow \infty}  \rho_{\rm 1-p} (x-x') \simeq \langle \phi(x) \rangle_0\,  \langle \phi(x')\rangle_0 \equiv n_0\,. \nonumber
\end{eqnarray}
The order parameter, playing the role of a macroscopic wave-function in condensed matter systems, is exactly the classical expectation value of the real scalar field 
\begin{eqnarray}
\langle \phi(x) \rangle_0 =\phi_\alpha(x)\,. \nonumber
\end{eqnarray}
As a main consequence, the density matrix of the system can be expressed as 
\begin{eqnarray}
\lim_{||x-x'|| \rightarrow \infty}  \rho_{\rm 1-p} (x-x') = \phi_\alpha(x)\, \phi_\alpha(x')\,, \nonumber
\end{eqnarray}
and thus we can identify the order parameter with 
\begin{eqnarray}
\phi_\alpha(x)= \sqrt{n_0} \, e^{\imath \theta(x)}\,. \nonumber
\end{eqnarray}
For a condensed matter system, the phase $\theta$ of the order parameter is usually a constant. Within the relativistic framework we are exploring, while the condensate still represents a coherent quantum state in which the $k_0$ mode has a macroscopic occupation, the phase turns out to be a function of the space-time point in order to be consistent with the Lorentz symmetry of space-time. 

This treatment is independent on the presence of an interaction term within the system, although its advantage is more evident when an interaction is present. Again, the comparison with the physics of very well understood condensed matter systems sheds light on this point. An ideal Bose condensate can be studied both in the Fock basis, which entails a fixed particle number representation, and in the coherent state basis. But for weakly interacting Bose gases, which have been studied by Bogolubov in the '40s, the analysis in terms of coherent states becomes crucial in solving the physical problem, for its many technical advantages. This is the case of the $\!\!\!\!\phantom{a}^4 {\rm He}$, for which interatomic interactions can not be disregarded. 

In what follows, we will argue that a similar procedure is worth to be extended to cosmological matter fields, for its unambiguity in recovering a consistent physical picture, and its versatility in performing calculations of power spectra with phenomenological interest.

\section{Cosmological perturbations: scalar fields} \label{scp}
\noindent 
How to define cosmological perturbation theory starting from the theory of quantum matter fields? We pursue results following few natural steps: i) associate the expectation values of quantum matter fields on coherent states to classical quantities in the standard framework; ii) identify those coherent states for bosonic and fermionic quantum matter fields, using known results in condensed matter and in representation theory; iii) recover classical perturbed fields in the standard picture by perturbing the coherent state in the relevant expectation values. Following this procedure, we will show that it is possible to define perturbation theory for objects of any statistics, and not only for scalar fields as commonly considered in the literature. 

In this section we start focusing on the case of real scalar field matter theories. Scalar fields, which are often encountered in several models of high energy physics, can easily help us to achieve a preliminary understanding of cosmological perturbations from the quantum point of view, to be deployed later to the case of physical matter fields.

We then move to our considerations taking into account canonical quantum scalar field $\hat{\phi}$, whose action is specifies by the assignment of the potential $V(\hat{\phi})$. Such a scalar field is governed by the action

\begin{align}
\label{action}
&S[\hat{\phi}]=\int d^4x\sqrt{-g}\bigg[\frac{1}{2}\partial_{\mu}\hat{\phi}\partial^{\mu}\hat{\phi}-V(\hat{\phi})\bigg]\,,
\end{align}
to which it corresponds the stress-energy tensor 

\begin{align}
\label{StressEnergyTensor}
&\hat{T}^\mu_\nu= \partial^\mu\hat{\phi}\partial_\nu\hat{\phi}-\delta^\mu_\nu \bigg[\frac{1}{2}\partial_{\lambda}\hat{\phi}\partial^{\lambda}\hat{\phi}-V(\hat{\phi})\bigg]\,.
\end{align}
The symmetries of the Friedmann-Lema\^itre-Robertson-Walker (FLRW) background, homogeneity and isotropy, imply that the scalar field will depend only on time and, hence, the resulting stress-energy tensor will be diagonal. Therefore, the energy density $\hat{\rho}$ and the pressure $\hat{p}$ associated with the scalar field simplify to
\begin{align}
\label{pressure}
\hat{\rho} &=\ \frac{1}{2}\dot{\hat{\phi}}^2+V(\hat{\phi}),\\
\hat{p} &=\ \frac{1}{2}\dot{\hat{\phi}}^2-V(\hat{\phi}).
\end{align}
The equation of motion for the quantum field $\hat{\phi}$ easily follows from the action \eqref{action}, and within the assumption of homogeneity and isotropy reads on the FLRW universe:
\begin{align} \label{eom}
\ddot{\hat{\phi}} + 3 H \dot{\hat{\phi}} + V'(\hat{\phi}) =0.
\end{align}
\\

In the ordinary theory of cosmological perturbation, which we have summarized for completeness in App.~III, all the operatorial quantities are substituted with their classical counterparts. The latter are then decomposed into their classical background components, which follow the dynamics of the FLRW background under scrutiny, plus perturbations, which are finally addressed as quantum fluctuations over the classical background. For instance, in the case of inflation quantum fluctuations act as the primordial seeds for the cosmological inhomogeneities. Applied to a single (homogeneous and isotropic) background scalar field, perturbation reads 
\begin{align}
\label{fi}
&\phi(x,t)=\bar{\phi}(t)+\delta\phi(x,t)\,.
\end{align}
Using the corresponding classical expression for the Lagrangian density \eqref{action}, and for its derived quantities \eqref{StressEnergyTensor}-\eqref{eom}, we can easily compute the expansions in the perturbation $\delta\phi$ of the pressure $p$, {\it i.e.} the perturbed pressure $\delta p$, and the perturbed energy density $\delta\rho$, together with the equation of motion for the perturbations variables $\delta\phi$, namely
\begin{align}
\label{deldenpre}
&\delta\rho=-\frac{1}{2}\dot{\bar{\phi}}\delta\dot{\phi}+V'(\bar{\phi)}\delta\phi\,,\\
&\delta p=-\frac{1}{2}\dot{\bar{\phi}}\delta\dot{\phi}-V'(\bar{\phi)}\delta\phi\,,\nonumber\\
\label{eomdle}
&\delta\ddot{\phi}+3H\delta\dot{\phi}-\frac{1}{a^2}\nabla^2\delta\phi +V''(\bar{\phi})\delta\phi=0\,.
\end{align}
Usually \eqref{eomdle} are solved (classically) by expanding the fluctuation $\delta\phi$ in complex exponentials with space-coordinates dependence that multiply time-dependent functions, since generically we are dealing with curved space-time:
\begin{align}
\label{furdf}
&\delta\phi(x,t)=\int \frac{d^3k}{(2\pi)^3} \big[\delta\phi_k(t)c_k e^{ikx}+\delta\phi^*_k(t)c^\dagger_k e^{-ikx}\big]\,,\\
\label{eomfur}
&\delta\ddot{\phi}_k+3H\delta\dot{\phi}_k+\frac{k^2}{a^2}\delta\phi_k +V''(\bar{\phi})\delta\phi_k=0\,.
\end{align}
Only at the end, within the standard procedure, one performs the canonical quantization by promoting the Fourier coefficients $c_k$ and $c_k^*$ to quantum creation and annihilation operators $\hat{c}_k$ and $\hat{c}^\dagger_k$ that fulfill commutation relations. \\

In our proposal, standard background fields $\phi$ are rather substituted with the expectation values of the quantum field $\hat{\phi}$ on coherent states $|\alpha\rangle$, namely $\phi_\alpha:=\langle \alpha| \hat{\phi}|\alpha\rangle$, and so forth for their functional that reproduce all the possible observable quantities, including the kinetic terms and the potential $V(\phi)$.   

To proceed with our analysis, we need to specify the potential $V(\phi)$, and then make some approximations in order to extract physical predictions, namely informations about the power spectrum of the CMB radiation. In what follows, we then focus on standard slow-roll inflation (see {\it e.g.} Appendix IV), and propose a prescription to associate quantum operators to perturbations. 

\subsection{Coherent states in perturbation theory}
\noindent 
There are several options of perturbed states that are worth to be explored when attempting to develop the theory of quantum perturbations from our perspective. Probably the naivest option would be to explore the perturbed state 
\begin{eqnarray} \label{sba}
\left|\alpha\right\rangle =\left|\alpha_{0}\right\rangle +\left|\alpha_{1}\right\rangle, 
\end{eqnarray}
in which both $|\alpha_{0}\rangle$ and $|\alpha_{1}\rangle$ in the right hand side of \eqref{sba} are coherent states, and an infinitesimal perturbation parameter is meant to multiply the second term. 
Nonetheless this option must be disregarded, since we are mainly interested in finding a coherent states that save the interpretation of semi-classiclity. Conversely, the state on the left hand side of \eqref{sba} is not a coherent state. 

We consider instead a coherent state labelled by a function $\alpha+\delta\alpha$, where $\delta\alpha$ is infinitesimal respect to $\alpha$
\begin{eqnarray}
\left|\alpha+\delta\alpha\right\rangle \,.
\end{eqnarray}
It is straightforward to notice that our definition of classical field is linear in the function that labels it
\begin{eqnarray}
\left\langle \alpha+\delta\alpha\right|\phi\left(x\right)\left|\alpha+\delta\alpha\right\rangle &=&\phi_{\alpha+\delta\alpha}\left(x\right) \\
&=&\phi_{\alpha}\left(x\right)+\phi_{\delta\alpha}\left(x\right) \nonumber\\
&=&\left\langle \alpha\right|\phi\left(x\right)\left|\alpha\right\rangle +\left\langle \delta\alpha\right|\phi\left(x\right)\left|\delta\alpha\right\rangle\,. \nonumber
\end{eqnarray}
For the same reason exposed above, we immediately understand that 
\begin{eqnarray}
\left|\alpha+\delta\alpha\right\rangle \neq\left|\alpha\right\rangle +\left|\delta\alpha\right\rangle\,. 
\end{eqnarray}
Thus we will not consider the superposition of two coherent states (which is not a coherent state), but instead a shifted coherent state. 

Furthermore, given any analytic function of the scalar field, we can trivially recover its classical limit:
\begin{eqnarray}
\left\langle \alpha+\delta\alpha\right|:V\left(\phi\right):\left|\alpha+\delta\alpha\right\rangle =V\left(\phi_{\alpha}+\phi_{\delta\alpha}\right)\nonumber\\
\approx  V\left(\phi_{\alpha}\right) +V'\left(\phi_{\alpha}\right)\phi_{\delta\alpha}+ \ldots.
\end{eqnarray}
Notice that perturbations of the label of the coherent state $|\alpha\rangle$ represents a perturbation in the total number density of the state, as defined in condensed matter, and summarized in \eqref{N}.  

Now that we have specified how to recover perturbed quantities in this framework, we can go back to the real scalar field action. The dynamics of the scalar field $\phi$, specified by the equation of motion \eqref{eom} that is in turn derived from the theory \eqref{action}, is cast at the operatorial level. Though we can extract the classical equation of motion out of \eqref{eom}, by simply taking its expectation value on the background state $|\alpha\rangle$, which represents the infrared matter wave functions of the Universe. Thus the statement about slow-roll must be now intended as a ``weak statement'', which concerns the condensate matter state that drives inflation but preserves background FLRW symmetries. As a matter of fact, from the expectation value of the operatorial equation it follows immediately  
\begin{eqnarray}
\langle \alpha |  \ddot{\widehat{\phi}}+ 3 H \dot{\hat{\phi}} + \widehat{V'(\phi)} | \alpha \rangle=0.
\end{eqnarray}
Ordinary slow-roll condition follows, when disregarding  $\langle \alpha |  \ddot{\widehat{\phi}} | \alpha \rangle$ with respect to the other terms:
\begin{eqnarray} \label{srb}
3 H \phi_\alpha \simeq - V(\phi_\alpha)\,.
\end{eqnarray}
This represents an approximated equality that is deployed in finding the perturbation variable, or in other words it represents the background value of operatorial quantities to be used while reshuffling the perturbed Einstein equations, namely
\begin{eqnarray} \label{peq}
\delta G_{\mu \nu}= \frac{8 \pi G}{c^4} \langle \alpha+\delta \alpha  |  \widehat{T_{\mu \nu}(\phi)}  | \alpha+\delta\alpha  \rangle\Big|_{\rm O(  \delta \alpha)}\,,
\end{eqnarray}
in which only the first order in $\delta \alpha$ is chosen in the second hand side of \eqref{peq}.  

Following the same steps of the standard derivation of the curvature perturbation variable (see Appendix IV), we obtain 
\begin{eqnarray} \label{sas1}
3(\zeta + \psi ) \langle \alpha | \widehat{\rho}+ \widehat{p} \,| \alpha \rangle = - \langle \alpha +\delta  \alpha | \widehat{\rho}| \alpha +\delta  \alpha \rangle\Big|_{\rm O(  \delta \alpha)}\,,
\end{eqnarray}
in which the second hand side is taken at $O(  \delta \alpha)$, and gravitational perturbation variables are assumed here to be classical quantities.

\subsection{New prescription for cosmological perturbations}
\noindent
We provide at this point a straightforward and natural prescription to recover  scalar cosmological perturbations theory starting from its quantum counterpart. The recipe amounts in recovering the semiclassical limit using the expectation value on the coherent states, identifying the perturbation, and than re-quantizing the perturbation using its newly derived equation of motion. 
If we identify the perturbation $\phi_{\delta\alpha}\equiv \delta \phi$ and the background field with $\bar\phi = \phi_\alpha$, we recover exactly the classical expressions that are used in the standard theory of cosmological perturbations, before quantizing $\delta \phi$. Notice that we can select $\alpha$ in order to reproduce any background configuration.

To be more specific, in order to derive the theory of scalar cosmological perturbations in this framework we wish to find a suitable curvature perturbation variable operator, such that its expectation value at $O(  \delta \alpha)$ on states $| \alpha +\delta  \alpha \rangle$ corresponds to the the value of the curvature perturbation variable $\zeta(t,x^i)$. Mimicking well known expressions, we can then proceed to the definition of an operator 
\begin{eqnarray} \label{sas2}
-\widehat{\Xi}= \widehat{1\!\!1}\, \psi(t, x^i)\, + \frac{\widehat{\rho}}{3 \langle \alpha| \widehat{\rho}+ \widehat{p} \, |\alpha\rangle}\,,
\end{eqnarray}
which enjoys this property. 

We can now apply the tools of our analysis to the relevant case of chaotic inflation, in which the potential is quadratic in the scalar field:
\begin{eqnarray}
\widehat{\rho}\simeq\widehat{V(\phi)}=\frac{1}{2} m^2 \widehat{\phi^2}\,.
\end{eqnarray} 
Its expectation value on perturbed states $| \alpha +\delta  \alpha \rangle$ reads
\begin{eqnarray}
&&\!\!\!\!\!\!\!\!\!\!\!\!\!\!\!\!\!\!\!\!\!\!\!\langle \alpha +\delta  \alpha |\, \widehat{\rho} \, | \alpha +\delta  \alpha \rangle= \nonumber \\
&=& \lim_{x\rightarrow y}\frac{1}{2} m^2 \, \langle \alpha +\delta  \alpha |  \,\widehat{\phi}(x) \, \widehat{\phi}(y)\, | \alpha +\delta  \alpha \rangle \nonumber\\
&=&  \lim_{x\rightarrow y}\frac{1}{2} m^2 \, \langle 0  | D( \alpha +\delta  \alpha)^\dagger 
\, \widehat{\phi}(x)\, 
D( \alpha +\delta  \alpha)\times\nonumber\\  && D( \alpha +\delta  \alpha)^\dagger 
\, \widehat{\phi}(y)\, 
D( \alpha +\delta  \alpha) \, |\alpha +\delta  \alpha \rangle \nonumber\\
&=& \frac{1}{2} m^2 \, [\phi_{\alpha +\delta  \alpha}(x)]^2 
\,,
\end{eqnarray} 
in which the standard regularization of the product of fields is intended, and which at $O(  \delta \alpha)$ becomes
\begin{eqnarray}
&&\langle \alpha +\delta  \alpha |\, \widehat{\rho} \, | \alpha +\delta  \alpha \rangle= m^2 \, \phi_\alpha(t) \, \delta \phi(x)\,, 
\end{eqnarray} 
having identified $\delta \phi=\frac{\delta \phi}{\delta \alpha}\delta \alpha$\,.

It is natural to identify the power spectrum of the scalar perturbations with the order $O(  \delta \alpha^2)$ of the expectation value of $\Xi$, namely
\begin{eqnarray} \label{pn}
\mathcal{P}_{\zeta} &=&  \lim_{x\rightarrow y}\, \langle \alpha +\delta  \alpha |\, \widehat{\Xi}(x)  \,\widehat{\Xi}(y) \, |\alpha +\delta  \alpha \rangle\Big|_{O(\delta \alpha^2)} 
\,,
\end{eqnarray} 
The ``background'' expectation values of $\widehat{\rho}$ and $\widehat{p}$ on $|\alpha \rangle$ within \eqref{pn} can be approximated in the standard way, under slow-roll approximation \eqref{srb}, and hence contributes to recreate the pre-factor $\frac{1}{m_{\rm Pl}^4} (\frac{V}{V'})^2$ in front of $\mathcal{P}_{\delta \phi}$. 

On the other hand, $\mathcal{P}_{\delta \phi}$ is still quadratic in $\delta \phi(x)$, the latter being now the solution for the operatorial equation of motion, evaluated at   (See Appendix IV). 

\section{Macroscopic states of matter: fermionic fields}

\noindent
The same formalism developed for bosonic fields and summarized in the previous sections can be extended to fermionic fields, generalizing their semiclassical applications to superconductors, as studied in the literature. Such an extension can not be achieved by means of the trivial definition of eigenstates of the annihilations operators for fermion particles and antiparticles. The reason is simple, and relies on the Pauli exclusion principle: single fermion states have occupation number 0 or 1, thus it is not possible to have a macroscopic number of fermions in a single plane-wave state. Mathematically, defining coherent states in such a way would involve the use of grassmannian numbers, and the resulting state would not be part of the physical Hilbert space. In this case the expectation values of the relevant fermionic bilinears, the observable operators entering the energy-momentum tensor in the Einstein equations, would not be real numbers, thus physically meaningless within the framework adopted to investigate the geometrodynamics of spacetime. There is indeed a crucial ingredient that must be taken into account: coherent macroscopic states must be developed in terms of pairs of fermions, which means that the role of photons or bosonic quanta is now played by pair of electrons or fermions.

Historically, R.~Schrieffer (see {\it e.g.} Ref.~\cite{BCS}) was the first one who managed to write a coherent many-particle wave-function for fermions while describing mathematically the ground state of superconducting atoms. He achieved this goal by deploying the understanding that Bardeen and Cooper realized about the binding of electrons in superconductor (see Ref.~\cite{BCS}). Schrieffer built indeed a macroscopic coherent state in which a very large number of pairs are all in the same state. In a BCS state \cite{BCS} electrons pairs must not be confused with bosons, as in stead it happened in the earlier theory by Schafroth, Blatt and Butler (see Refs.~\cite{SBB, S}) of superconductivity seen as a Bose condensate of electron pairs. For a BCS state each electron take part in the pairing, and this is experimentally confirmed by data on superconductivity. Nonetheless, at high critical temperature, pairs do not have a large overlap, and condensation {\it \`a la } Schafroth, Blatt and Butler may arise \cite{S, SBB}. We shall not be concerned anyway with this peculiar situation, in which the pairs form a ``pseudomolecule'' whose size is much smaller than the average distance between them. This system, despite having properties similar to those of a charged Bose-Einstein gas, including Meissner effect and critical temperature of condensation \cite{SBB3}, can be rather accounted for in the most general framework of the BCS states. The former can be indeed regraded as ``bipolarons'', {\it i.e.} localized spatially nonoverlapping Cooper pairs that  form by strong electron-phonon interaction \cite{AR}. 

In the cosmological framework we aim at developing, semiclassical states that belong to the fermionic Fock space fall naturally in the class of states of physical interest well described by BCS states. In other words, the states to be considered in the cosmological set-up have closer analogies with non-overlapping Cooper pairs in the weak-interacting regime than with the states that have been advocated while implementing the concept of pseudomolecules. We shall then proceed to develop BCS states for fermion matter fields in cosmology.

\subsection{BCS coherent state and Bosonization}
\noindent
Bosonization consists in replacing a fermionic system, and the related Hilbert space, with a bosonic theory, completely equivalent from the physical point of view in that it encodes identical spectra and interactions. The procedure turns out to be particularly advantageous to the analysis of fermionic systems, given that many powerful techniques developed for bosonic systems can be now deployed while describing  fermionic systems. Nonetheless, a major limitation of bosonization consists in the fact that this can be naturally achieved only in one space dimension \cite{Bosonization1}. 

It is possible to define a bosonization procedure in any dimension, using a completely antisymmetric gauge field of rank $space\text{-}dimension-1$, which is then dual to a scalar field. Although the bosonic theory we are led to in this way is guaranteed to exist, it is not required to have many of the usual properties that we tend to take for granted in the one-space-dimension case, such as locality \cite{Bosonization2}.

What we are trying to address here is different, we are not requiring to map the fermionic Hilbert space into a bosonic equivalent, but we want rather to find a (set of) state(s) with the correct macroscopic coherent behavior.

\subsection{Exploring our quasi particle options}
\noindent
In analogy to the bosonic case we then consider free fermionic fields on flat (Minkowski) background, whose density Lagrangian and Hamiltonian in natural units read respectively 
\begin{eqnarray}
L(x)&=& \overline{\psi}(x) i \gamma_\mu \partial^\mu \psi(x)\,,\\ 
H(x)&=& \overline{\psi}(x) (- i \gamma_i \partial^i ) \psi(x)\,. \nonumber
\end{eqnarray}
The field $\psi(x)$ (and similarly its conjugated momentum) is decomposed in the Fock basis of the harmonic linear oscillators, as a superposition of creation and annihilation operators for each mode: 
\begin{eqnarray}
\psi\left(x\right)=\int\displaylimits_k
\sum_{s=\pm}\left(a_{k}^{s}u^{s}\left(k\right)e^{-ikx}+b_{k}^{\dagger s}v^{s}\left(k\right)e^{+ikx}\right)\,,
\end{eqnarray}
where $u$ and $v$ are the particle and anti-particle wave functions, $a$ and $b$ are the annihilation operators of the fermion and the anti-fermion, and the integral over the momenta has the proper measure invariant under the action of the space-time isometries.

We will introduce here some Cooper pair-like creation (annihilation) operators. These operators do not obey normal Bose commutation laws, and so they cannot be regarded as creating or destroying boson particles.

We will look for a uniform translationally invariant solution, and so it is more convenient to work in k space. Let us define the pair creation operator by,
\begin{eqnarray}
\label{cooperpair}
c^{\dag}_k=a^{\dag}_{k\uparrow} b^{\dag}_{-k\downarrow}\,, \qquad c_k= b_{-k\downarrow} a_{k\uparrow}\,. 
\end{eqnarray}
Note that this pair operators have the following commutation relations:
\begin{align}
\label{comrelc}
\Big[c_k,c_{k'}\Big]=0\,, \qquad \left[c^\dag_k,c^\dag_{k'}\right]= 0\,,\\
\left[c_k,c^{\dag}_{k'}\right]=(1-b^\dag_{-k\downarrow}b_{-k\downarrow}-a^\dag_{k\uparrow}a_{k\uparrow})\delta_{k,k'}\, .
\end{align}
Moreover the pair creation operator is idempotent
\begin{equation}
c^\dag_k c^\dag_k = a^{\dag}_{k\uparrow} b^{\dag}_{-k\downarrow} a^{\dag}_{k\uparrow} b^{\dag}_{-k\downarrow} = 0\,,
\end{equation}
since it contains two identical fermion creation operators. In terms of this operator, following the usual construction for BCS theory, we propose a coherent state labelled by the function $\alpha(k)\,:\,\mathbb{R}^{3}\to\mathbb{C}$
\begin{equation}
\label{bcs-like-cs}
\left|\alpha\right\rangle \equiv e^{\int d^{3}k\,\alpha\left(k\right)c_{k}^{\dagger} - \alpha^*\left(k\right) c_k}\left|0\right\rangle\, = D\left(\alpha\right) \left|0\right\rangle\,.
\end{equation}
As for the bosonic case, the displacement operator $D$ is unitary $D^{\dagger}\left(\alpha\right) D\left(\alpha\right)=1$, and generates the coherent state $\left|\alpha\right\rangle $ from the vacuum.

\subsection{BCS-like state and $SU(2)$ coherent states}
\noindent
In analogy with the construction mentioned in Section \ref{sh} we will now explore the relation between the BCS-like state introduced in the previous sections and the group coherent states of $SU(2)$. 
Consider the Hilbert space generated by a couple of \textit{fermionic} creation and annihilation operators $a$ and $b$, such that
\begin{equation} \nonumber 
\left\lbrace a , a^\dagger \right\rbrace = \left\lbrace b , b^\dagger \right\rbrace = 1\,, \qquad \left\lbrace a , b \right\rbrace =\left\lbrace a , a \right\rbrace =\left\lbrace b , b \right\rbrace = 0\,.
\end{equation}
Using these operators we can define the generators of a $SU(2)$ algebra, which read explicitly
\begin{align}
J_{1} & =\frac{1}{2}\left(a^{\dagger}b^{\dagger}+h.c.\right)\ ,\\
J_{2} & =-\frac{i}{2}\left(a^{\dagger}b^{\dagger}-h.c.\right)\ ,\\
J_{3} & =\frac{1}{2}\left(a^{\dagger}a+b^{\dagger}b-1\right)\ .
\end{align}
These operators are clearly hermitian and obey the following commutation rules
\begin{equation}
\left[J_{i},\ J_{j}\right]=i\epsilon_{ij}^{\phantom{{ij}}k}J_{k}\ .
\end{equation}
In terms of the creation and annihilation operators the Casimir and the ladder operators of the algebra read
\begin{align*}
J^{2} & =-\frac{3}{4}\left(2a^{\dagger}b^{\dagger}ab+b^{\dagger}b+a^{\dagger}a-1\right)\,,\\
J^{+} & =J_{1}+iJ_{2}=a^{\dagger}b^{\dagger} \,, \qquad J^{-}  =J_{1}-iJ_{2}=ba\,.
\end{align*}
Notice how the $J^\pm$ operators can be interpreted as the Cooper pair-like creation and annihilation operators \eqref{cooperpair}.\\
We can than consider generalized coherent state of this $SU(2)$ group (for more details see Appendix II) labelled by a unitary vector $\hat n$, or equivalently by a complex number 
\begin{eqnarray}
\xi = \frac{n_x + i n_y}{1+n_z}
\end{eqnarray}
entering the reshuffled expression 
\begin{eqnarray}
\label{SU2-complex-cs}
\left|\hat{n}\right\rangle &=&D(\hat n)\left|j,-j\right\rangle = \nonumber \\
\left|\xi\right\rangle &=&\exp\left(-\frac{\bar{\xi}}{|\xi|}\arctan(|\xi|) J^{+}+\frac{\xi}{|\xi|}\arctan(|\xi|)J^{-}\right) \nonumber \\
&\phantom{a}& \times \left|j,-j\right\rangle\,.
\end{eqnarray}\\

We can ask ourselves in which $SU(2)$ irreducible representation transforms the fermionic vacuum. The answer is straightforward, if we consider that
\begin{align*}
J^{2}\left|0\right\rangle  & =-\frac{3}{4}\left(2a^{\dagger}b^{\dagger}ab+b^{\dagger}b+a^{\dagger}a-1\right)\left|0\right\rangle =\frac{3}{4}\left|0\right\rangle \,,\\
J_{3}\left|0\right\rangle  & =\frac{1}{2}\left(a^{\dagger}a+b^{\dagger}b-1\right)\left|0\right\rangle =-\frac{1}{2}\left|0\right\rangle \,.
\end{align*}
Thus the fermionic vacuum corresponds to the lowest weight state in the $j=1/2$ $SU(2)$ irreducible representation.
By comparing \eqref{bcs-like-cs} and \eqref{SU2-complex-cs} the BCS-like coherent state can be interpreted as the tensor product of $SU(2)$ coherent states (one for each Cooper-pair).
This correspondence turns to be extremely useful while studying the semiclassical properties of the fermionic bilinears. To address this point, let us start considering the following expression first
\begin{equation}
\label{Iexpression}
I=\int_{k}a_{k\uparrow}^{\dagger}b_{-k\downarrow}^{\dagger}A_k+b_{-k\downarrow}a_{k\uparrow}A^*_k+a_{k\uparrow}^{\dagger}a_{k\uparrow}B_k-b_{-k\downarrow}b_{-k\downarrow}^{\dagger}B_k\ ,
\end{equation}
where $A_k$ and $B_k$ are numbers, possibly $k$ dependent. Notice that the requirement of $I$ to be hermitian fix $B_k$ to be real. Notice that it is possible to rewrite $I$ in terms of $\vec{J}_k$, the $SU(2)$ generators corresponding to the fermionic creation and annihilation operators $a_{k\uparrow}$ and $b_{-k\downarrow}$, {\it i.e.}
\begin{equation} \label{f1}
I =\int_{k} 2 A_k J^+_{k} + 2 A^*_k J^-_{k} +2 B_k J_{3k} = \int_{k} \vec{\zeta_k} \cdot \vec{J}_{k}\,,
\end{equation}
where we have defined for convenience $\vec{\zeta_k}= \left(2 \mathrm{Re}(A_k),\,2\mathrm{Im}(A_k),\,2B_k\right)$. The expectation value of such expression on a coherent state is easily computed, using the properties of the coherent state.
\begin{equation} \label{f2}
\left\langle \hat{n}\right| I \left|\hat{n}\right\rangle = \int_{k} \vec{\zeta_k} \cdot \left\langle \hat{n}\right| \vec{J}_{k} \left|\hat{n}\right\rangle = \int_{k} \vec{\zeta_k} \cdot \hat{n}_k\,.
\end{equation}
Now, the expectation value of the fermionic bilinear on a coherent state is equivalent to the expectation value of the operator $I$:
\begin{equation} \label{f3}
\left\langle\hat{n}\right|\bar{\psi}\left(x\right)\psi\left(x\right)\left|\hat{n}\right\rangle = \left\langle\hat{n}\right|I\left|\hat{n}\right\rangle = \int_{k} \vec{\zeta_k} \cdot \hat{n}_k\,.
\end{equation}
We can look explicitly at the Fourier expansion of the bilinear
\begin{align}
\bar{\psi}\left(x\right)\psi\left(x\right)=\sum_{st=\pm}\int\displaylimits_{k_1,k_2}&a_{k_{1}}^{\dagger s}b_{k_{2}}^{\dagger t}\bar{u}^{s}\left(k_{1}\right)v^{t}\left(k_{2}\right)e^{+i\left(k_{1}+k_{2}\right)x}\nonumber\\
&\hspace{-.4cm}+b_{k_{1}}^{s}a_{k_{2}}^{t}\bar{v}^{s}\left(k_{1}\right)u^{t}\left(k_{2}\right)e^{-i\left(k_{1}+k_{2}\right)x}\nonumber\\
&\hspace{-.4cm}+a_{k_{1}}^{\dagger s}a_{k_{2}}^{t}\bar{u}^{s}\left(k_{1}\right)u^{t}\left(k_{2}\right)e^{+i\left(k_{1}-k_{2}\right)x}\nonumber\\
&\hspace{-.4cm}+b_{k_{1}}^{s\dagger} b^{t}_{k_{2}}\bar{v}^{s}\left(k_{1}\right)v^{t}\left(k_{2}\right)e^{-i\left(k_{1}-k_{2}\right)x} \nonumber
\end{align}
and recognize that the expectation value on a coherent state gets non-vanishing contribution only if $k_2=-k_1$ and spins are opposite by construction. If we identify $$A_k = \bar{u}^{+}\left(k\right)v^{-}\left(-k\right)$$ and $$B_k= \bar{u}^{+}\left(k\right)u^{+}\left(k\right)=-\bar{v}^{-}\left(-k\right)v^{-}\left(-k\right),$$ we recover exactly the expression \eqref{Iexpression}.
Notice that other kind of bilinear has exactly the same $SU(2)$ structure of the scalar bilinear considered above, the main difference residing in the explicit form of $\vec{\zeta_k}$. The latter is indeed a vector in the internal indices space that can also acquire space-time indices, and can eventually transform as an axial (internal) vector under parity transformations, in order to reproduce the Fourier transform of a generic element of the fermionic bilinears' Clifford algebra.

In analogy with the bilinear also the quadrilinear expectation values can be expressed in terms of $SU(2)$ generators, namely
\begin{align*}
&\left\langle\hat{n}\right|\bar{\psi}\left(x\right)\psi\left(x\right)\bar{\psi}\left(x\right)\psi\left(x\right)\left|\hat{n}\right\rangle \\
&\qquad= \left\langle\hat{n}\right| \int\displaylimits_{k,k'} \vec{J}_{k'}\cdot M_{kk'} \cdot \vec{J}_{k}\left|\hat{n}\right\rangle \\
&\qquad= \int\displaylimits_{k,k'} \hat{n}_{k'}\cdot M_{kk'} \cdot \hat{n}_k\,.
\end{align*}
\\

It will be useful for the study of cosmological perturbations that follows in the next section to explore the behavior of the generalized coherent state under infinitesimal variation of the label. We \textit{define} the perturbed coherent state labelled by a vector $ \hat{n}$ and the perturbation $\delta \hat{n}$ as the one obtained by the subsequent action of the appropriate displacement operators
\begin{equation}
\left| \hat{n} , \delta \hat{n} \right\rangle \equiv  D\left(\delta \hat{n} \right) D\left( \hat{n} \right) \left| 0 \right\rangle = \left|R\left(\hat{z},\delta \hat{n}\right) \hat{n} \right\rangle\,,
\end{equation}
where $R\left(\hat{z},\delta \hat{n}\right)$ is the rotation matrix that transforms $\hat{z}$ (the north pole direction) into $\delta \hat{n}$, and the last equality is valid up to an irrelevant phase. If then $\delta \hat{n}$ is infinitesimal, the rotation matrix $R\left(\hat{z},\delta \hat{n}\right)$ is almost the identity:
\begin{equation}
\left| \hat{n} , \delta \hat{n} \right\rangle \approx \left| \hat{n} + \delta \hat{n}\times\hat{n} \right\rangle\,.
\end{equation}
Notice also that $|| \hat{n} + \delta \hat{n}\times\hat{n}||=1 + O(\delta \hat{n})$, so to identify a good coherent state. The expectation value of the bilinear on the perturbed state is the easily found to be 
\begin{eqnarray} \label{f4}
\left\langle \hat{n} , \delta \hat{n} \right|\bar{\psi}\left(x\right)\psi\left(x\right)\left|\hat{n} , \delta \hat{n}\right\rangle &=& \int_{k} \vec{\zeta_k} \cdot \hat{n} + \int_{k}\vec{\zeta_k} \cdot \delta \hat{n}_k\times\hat{n}_k=\nonumber \\
= \left\langle \hat{n}  \right|\bar{\psi}\left(x\right)\psi\left(x\right)\left|\hat{n} \right\rangle &+&  \left\langle \delta \hat{n} \right| \int_{k}\hat{n}_k \times \vec{\zeta_k}\cdot  \vec{J}_k \left| \delta \hat{n} \right\rangle \,,
\end{eqnarray}
which is linear in the perturbation parameter. For its clear physical implications, this result is certainly the most important that we have derived in this section. We will dwell more on it in what follows. 

\section{Cosmological perturbations: fermionic fields}
\noindent 
The usual ``no-go argument'' against linear cosmological perturbations obtained in terms of fermionic Dirac fields is the following: 
\begin{enumerate}
\item
if one takes into account linear perturbations of bilinear (in the Dirac fields) operators, treating fermionic fields as classical these will read
\begin{eqnarray}
\delta(\overline{\psi} \Gamma^{\aleph}\psi)=\delta\overline{\psi} \, \Gamma^{\aleph}\psi+ \overline{\psi} \Gamma^{\aleph}\, \delta\psi\,,
\end{eqnarray}
with $\aleph=1,\dots 16$ an index that labels the elements of the Clifford algebra. For simplicity, let us fix $ \Gamma= 1\!\!1$ in the internal space of the Dirac fields, and deal with the quantity we have denoted with $I$ in the previous section; 
\item
on a FLRW space-time background, we treat $\delta \psi(x)$ as perturbed fields, with a dependence on the space-time point, and $\psi(t)$ as background fields, which only depend on some cosmological time $t$;
\item 
we quantize both the fields, resorting eventually to two different Hilbert spaces. For the background part $\psi(t)$ we imagine to use an Hilbert space whose vacuum shares the same symmetries of the background FLRW metric (in particular, if we are dealing with a de Sitter background, the vacuum state will turn out to be the Bunch-Davies vacuum \cite{Bunch:1978yq}, enjoying de Sitter symmetries); 

\item 
we recognize that on any state $|\gamma\rangle$ compatible with the symmetries of the background, the expectation values of $\psi(t)$ would be zero. Indeed, rotating the $\psi$ about the cartesian axis $z$ of an angle $2\pi$, by means of a SU(2) group element $R(\hat{z}, \hat{z})\equiv R$, and using the invariance under rotation of the state $| \gamma \rangle$, we easily find that
\begin{eqnarray}
\langle \gamma | \psi | \gamma \rangle= \langle \gamma | R^\dagger \psi R | \gamma \rangle= - \langle \gamma | \psi | \gamma \rangle \,, 
\end{eqnarray}
from which it follows $\psi(t)=\langle \gamma | \psi | \gamma\rangle=0$. Notice however that one could consider in stead of $|\gamma\rangle$ a state with non definite spin, which does not transform trivially under spin rotation. This is the state considered in \cite{Endlich:2013spa}, which invalidates the argument above. Nonetheless, we notice that such a state is not suitable to define a semiclassical limit, and thus a background field $\psi(t)$.

\end{enumerate}

\noindent
This argument is commonly advocated to show that linear perturbations of fermionic bilinear are not possible. We argue that this argument is incorrect, for several reasons. 
First of all, Dirac fields are operators that are subjected to anti-commutative relations, thus can not be simply treated as classical fields. Second, we can not consider the expectation value of a fermion field, because this would not be a proper observable: it does not fulfill the requirement of micro-causality, and is not ``gauge invariant'' in the particle physics sense, namely is not invariant under SL(2,$\mathbb{C}$) transformations (for the same reasons, a single fermion field can not transform covariantly under space-time transformations). Third, the procedure of quantizing a background field and separately a perturbed field poses some ambiguities at the level of the definition of the Hilbert space of the fermionic theory. Are the two Hilbert spaces different? If they are not different, in which sense the field operator is a perturbed quantity?
Indeed, the expectation value of the product of two operators is not equal to the product of two expectation values, unless the Hilbert spaces on which the two operators act are different. 

Below we propose a simple way to overcome all these unnecessary complications, and to avoid the inconsistencies related to the approach criticized above. The choice is actually natural: we just need to extend the treatment already outlined in Sec.~\ref{scp}, which was tailored for a theory of fields subjected to bosonic quantization, in order to account as well for fermion fields perturbations.  The way to achieve this result is also straightforward. Micro-causality and covariance under space-time transformations force us to write the main physical equations as function of observable bilinears $\mathcal{O}^\aleph$, or eventually their (regularized) products. 

Thus, following the same line of thought reported in the previous sections, the perturbed Einstein read
\begin{eqnarray} \label{peqf}
\delta G_{\mu \nu}= \frac{8 \pi G}{c^4} \langle \alpha+\delta \alpha  |  \widehat{T_{\mu \nu}(\mathcal{O}^\aleph)}  | \alpha+\delta\alpha  \rangle\Big|_{\rm O(  \delta \alpha)}\,,
\end{eqnarray}
in which now $|\alpha\rangle$ and $|\alpha+\delta \alpha \rangle$ are the BCS-like states defined in \eqref{bcs-like-cs}, in the fermionic Fock space. Perturbations analysis then follows the same steps as in \eqref{sas1} and \eqref{sas2}, provided that we recognize that 
\begin{eqnarray}
\rho=\rho(\mathcal{O}^\aleph)\,, \qquad p=p(\mathcal{O}^\aleph)\,,
\end{eqnarray}
and that perturbations of the fermion bilinears $\mathcal{O}^\aleph$ are achieved as in \eqref{f1}-\eqref{f3} and \eqref{f4}.

\section{Coherent states and gauge transformations }
\noindent
Before spelling out the conclusions, it is necessary to derive the transformation rules for the coherent states $|\alpha\rangle$ introduced so far. At this purposes, we first derive the transformation properties of the ladder operators. For the sake of clarity, we start the analysis with a straightforward case: space-time translations acting on a scalar field on Minkowski space-time. 

A real scalar field $\hat{\phi}(x)$ on flat space-time is expanded as in \eqref{scafi}. Invariance under space-time translations $x\rightarrow x'=x+\delta$ implies
\begin{eqnarray}
\hat{\phi}'(x')&=&e^{-i \hat{P}_\mu \delta^\mu }\hat{\phi}\left(x\right) e^{i \hat{P}_\mu \delta^\mu }=\hat{\phi}(x+\delta)\\
&=&\int_k
\left( e^{-i k \delta } \hat{a}_{k}e^{-i kx}+  e^{+i k \delta } \hat{a}_{k}^{\dagger}e^{+i k x}\right)\,, \nonumber
\end{eqnarray}
in which we have introduced the generators $\hat{P}_\mu$ of the abelian algebra $\mathcal{T}_4$ of space-time translations on Minkowski space-time, and which corresponds to a transformation on the ladder operators 
\begin{eqnarray} \label{trala}
\hat{a}_{k} \rightarrow e^{+i k \delta } \hat{a}_{k}\,, \qquad \hat{a}_{k}^{\dagger} \rightarrow e^{+i k \delta } \hat{a}_{k}^{\dagger}\,.
\end{eqnarray}
Since this property holds at the operatorial level, it must hold also as a weak property, on the expectation values $\langle \phi\rangle_\alpha$. This implies that $|\alpha\rangle$ must be invariant under space-time translations, if we are working in the Heisenberg picture in which the ladder operators must fulfill \eqref{trala} in order $\hat{\phi}$ to be invariant under space-time translations. Thanks to the invariance of the integration measure on the Fourier modes \cite{Weinberg_book}, a similar argument applies also to Lorentz transformations, provided that for those latter $x \rightarrow x'=\Lambda x$, and
\begin{eqnarray} \label{trala2}
\hat{a}_{k} \rightarrow  \hat{a}_{\Lambda^{-1} k}\,, \qquad \hat{a}_{k}^{\dagger} \rightarrow  \hat{a}^\dagger_{\Lambda^{-1} k}\,.
\end{eqnarray}
\\

While it is convenient to implement the same strategy when accounting for diffeomorphisms, we must anyway resort to a different analysis of the transformations, focusing on the Fourier parameters space in order to avoid referring to finite space-time transformations. As renown, this can not be implemented using the ordinary tools of Lie groups, as it happens instead for the case of Poincar\'e transformations. 

For simplicity, let us still consider to be on flat Minkowski space-time. We may think at space-time diffeomorphisms to be generated by infinitesimal vectors $\xi^\mu(x)$ through  
\begin{eqnarray}
x^\mu\rightarrow x'^{\mu}=x^\mu+\xi^\mu(x)\,,
\end{eqnarray}
to be formally implemented by the action of an element $\eta^{\mu}(\hat{P}) \in U(\mathcal{T}_4)$, belonging to the enveloping algebra of $\mathcal{T}_4$, and of infinitesimal order. The vector $\eta^\mu$ acts on the Fourier basis as
\begin{eqnarray}
\eta^\mu(\hat{P}) e^{-i kx}=\eta^\mu(k) e^{-i kx}\,,
\end{eqnarray}
and by definition generates space-time diffeomorphisms, by acting on the Fourier space as
\begin{eqnarray}\label{keta}
k^\mu\rightarrow k'^\mu=k^\mu+\eta^\mu(k)\,.
\end{eqnarray}
Relation \eqref{keta} will finally induce a transformation on the ladder operators 
\begin{eqnarray} \label{trala3}
\hat{a}_{k} \rightarrow  \hat{a}_{\eta(k)}\,, \qquad \hat{a}_{k}^{\dagger} \rightarrow  \hat{a}^\dagger_{\eta(k)}\,,
\end{eqnarray}
which is required in order to ensure the invariance of $\hat{\phi}(x)$ under diffeomorphisms. \\

We can now go back to our initial question: how does a coherent state transform under a generic change of coordinate $x\to y\left(x\right)$? Clearly we can Fourier expand the field in terms of plane wave in the new coordinates $y(x)$, namely 
\begin{eqnarray}
\hat{\phi}(y)=\hat{\phi}\left(y(x)\right)&=&\int_k
\left(\hat{a}_{k}e^{-i k y(x)}+  \hat{a}_{k}^{\dagger}e^{+i k y(x)}\right)\,, \nonumber
\end{eqnarray}
but also in terms of plane waves of the old coordinates $x$, using different ladder operators, {\it i.e.}
\begin{eqnarray}
\hat{\phi}\left(y(x)\right)=\widehat{\phi\circ y}\left(x\right)&=&\int_k
\left(\hat{\tilde a}_{k}e^{-i k x}+  \hat{\tilde a}_{k}^{\dagger}e^{+i k x}\right)\,. \nonumber
\end{eqnarray}
It is possible to show that the new ladder operators can be written as a linear combination of the old one \cite{Hawking} (they are different bases of operators that generate the same Hilbert space): 
\begin{equation}
\hat{\tilde a}_{k} = \int_{k'} \left(A(k,k')\hat{a}_{k'}+B(k,k')\hat{a}_{k'}^{\dagger}\right)\,,
\end{equation}
where $A$s and $B$s are complex coefficients that can be determined in terms of the Fourier transform of $e^{-i k y(x)}$ using the normalization condition $|A(k,k')|^2 - |B(k,k')|^2=\delta_{k,k'}$ (the latter property is obtained requiring that $\hat{\tilde a}_{k}$ and $\hat{\tilde a}^\dagger_{k}$ satisfies canonical commutation relation). Then a general coherent state can be written in terms of both the bases
\begin{eqnarray}
\left|\alpha\right\rangle=D\left(\alpha\right)\left|0\right\rangle &=& e^{\int d^{3}k\,\alpha\left(k\right)\tilde a_{k}^{\dagger}-\alpha^{*}\left(k\right)\tilde a_{k}}\left|0\right\rangle\\
&=& e^{\int d^{3}k\,\gamma\left(k\right)a_{k}^{\dagger}-\gamma^{*}\left(k\right) a_{k}}\left|0\right\rangle
\end{eqnarray}
where $\gamma(k)=\int_{k'} \left[ B^*(k,k')\alpha(k')-A(k,k')\alpha^*(k')\right]$. We can then conclude that under a general coordinate transformation a coherent state is mapped into another coherent state with a label that is the Bogolubov transform of the old label.

\section{Conclusions} 
\noindent 
We have shown that semi-classicality in cosmological frameworks allow to tackle issues and severe restrictions that otherwise might arise in the theory of cosmological perturbations.
Among many phenomenological consequences that we expect our analysis can offer, we focused in particular on the possibility of studying cosmological perturbations induced by fermionic fields at the linear order. We actually showed that following our procedure, cosmological perturbations that might arise due to fermionic fields can not be claimed to be vanishing a priori. 

Phenomenological consequences of the existence of such a macroscopic condensate state of matter follow, including the possibility of generating cross-correlation spectra directly from fermion perturbations. Several studies are in preparation to show the instantiation of this proposal within both the inflation and matter-bounce scenarios \cite{DMTV}, adapting to this procedure previous preliminary investigations on the phenomenological applications of the Dirac theory in cosmology \cite{ArmendarizPicon:2003qk, Cai:2008gk, Alexander:2014eva, Alexander:2014uaa}.

Fermion matter fields are ubiquitous in our current understanding of physics, both in the branches of particle physics and condensed matter. Especially in the field of condensed matter, the semiclassical limit of fermion matter fields has reached amazing theoretical and experimental results, and has faced what in the field of particle physics there was no need to address: the semiclassical limit. 

We acknowledge that our main inspiration, as researchers trained in the field of high energy physics, actually came from constructions developed in a different field as ours. We now believe that this cross-fertilization will be at the origin of novel conquests not only in theoretical physics, but in its very phenomenological related applications. Forthcoming studies \cite{BhramaDonaMarciano} will make clear what we expect to derive by following this line of thought.

\section*{Appendices}
\subsection*{Appendix I. Harmonic oscillator coherent state}
\noindent
We review below basic facts concerning coherent states for the harmonic oscillator, which is at the base of the definition of coherent states for quantum systems enjoying Bose-Einstein statistics. An harmonic oscillator coherent state $\left|\alpha\right\rangle $ is defined as the eigenstate of the annihilation operator $a$, with eigenvalues $\alpha\in\mathbb{C}$:
\begin{eqnarray}
a\left|\alpha\right\rangle =\alpha\left|\alpha\right\rangle \,.
\end{eqnarray}
Since $a$ is a non-hermitian operator the eigenvalue $\alpha$ is a complex number. Coherent states are characterized by the properties:
\begin{itemize}
\item the vacuum is a coherent state with $\alpha=0$;
\item the mean energy is $\left\langle \alpha\right|H\left|\alpha\right\rangle =\hbar\omega\left\langle \alpha\right|a^{\dagger}a+\frac{1}{2}\left|\alpha\right\rangle =\hbar\omega\left(\left|\alpha\right|^{2}+\frac{1}{2}\right)$;
\item the displacement operator can be defined,
\begin{eqnarray}
\label{HOdisplacement}
D\left(\alpha\right)=e^{\alpha a^{\dagger}-\alpha^{*}a}\,,
\end{eqnarray}
where $\alpha\in\mathbb{C}$ and $a$, $a^{\dagger}$ are the annihilation and creation operators. It is unitary $D^{\dagger}D=1$ and generates the coherent state $\left|\alpha\right\rangle $ from the vacuum $\left|0\right\rangle$, 
\begin{eqnarray}
\left|\alpha\right\rangle =D\left(\alpha\right)\left|0\right\rangle\,; 
\end{eqnarray}
\item the action of the displacement operator on the creation or annihilation operator displace them
\begin{align} \label{btri2}
& D\left(\alpha\right)^\dagger a D\left(\alpha\right) = a + \alpha\, ,\\
& D\left(\alpha\right)^\dagger a^\dagger D\left(\alpha\right) = a^\dagger + \alpha^* \, \nonumber;
\end{align}
\item the coherent state can be expanded on the Fock basis
\begin{eqnarray}
\left|\alpha\right\rangle =e^{-\frac{\left|\alpha\right|^{2}}{2}}\sum_{n=0}^{\infty}\frac{\alpha^{n}}{\sqrt{n!}}\left|n\right\rangle =e^{-\frac{\left|\alpha\right|^{2}}{2}}\sum_{n=0}^{\infty}\frac{\left(\alpha a^{\dagger}\right)^{n}}{n!}\left|0\right\rangle \,;
\end{eqnarray}
\item the scalar product of two coherent states reads $\left\langle \beta|\alpha\right\rangle =e^{-\frac{\left|\alpha\right|^{2}}{2}}e^{-\frac{\left|\beta\right|^{2}}{2}}e^{\alpha\beta^{*}}$ and $\left|\left\langle \beta|\alpha\right\rangle \right|^{2}=e^{-\left|\alpha-\beta\right|^{2}}$;
\item although the coherent states are not orthogonal, they form an over-complete set of states
\begin{eqnarray}
\frac{1}{\pi}\int d^{2}\alpha\left|\alpha\right\rangle \left\langle \alpha\right|=1\,.
\end{eqnarray}
\end{itemize}
An essential feature of these states is that their number uncertainty is proportional to the square root of the expectation value of the number operator $N=a^\dagger a$ on these states, since
\begin{eqnarray}
\langle  N \rangle\equiv \langle \alpha|  N | \alpha \rangle =|\alpha|^2 \,, \qquad \Delta N=\sqrt{\langle  N^2  \rangle-\langle  N \rangle^2}=|\alpha|\,, \nonumber
\end{eqnarray}
from which it follows that on a coherent state
\begin{eqnarray}
\frac{\Delta N}{\langle  N \rangle}\sim \frac{1}{\sqrt{\langle  N \rangle}}\,.
\end{eqnarray}
Approximation of many operator expectation values by mean-field values the follows through the replacement $N\simeq \langle  N \rangle$. Although $\langle  N \rangle$ is not a definite quantum number for the coherent states, being $\Delta N=\sqrt{\langle  N \rangle}$, nonetheless they posses a definite phase $\theta$. Since coherent states are defined for any $\alpha\in\mathbb{C}$ by means of $\alpha=|\alpha|e^{\imath \theta}$, one the operator $\theta$ can be introduced such that 
\begin{eqnarray}
\frac{1}{\imath} \frac{\partial}{\partial \theta}|\alpha\rangle= \langle  N \rangle |\alpha\rangle
\,.
\end{eqnarray}
This entails to recast coherent states in terms of the conjugated operators $N$ and $\theta$, for which the uncertainty principle can be recast as 
\begin{eqnarray}
\Delta N \Delta \theta\geq \frac{1}{2}\,.
\end{eqnarray}
While the energy eigenstates have a well defined $N$ but an arbitrary phase, coherent states do not carry definite values of number operator $N$, but are rather endowed with a fixed phase.

\subsection*{Appendix II. $SU(2)$ coherent states}
\noindent 
In this section we will briefly summarize definitions and properties of the $SU(2)$ generalized coherent state as defined in \cite{Perelomov}. \\
Given an generic $SU(2)$ element in some representation we can express it using the exponential map in terms of the generators of the corresponding $SU(2)$ algebra 
\begin{equation}
\left[J_{i},\ J_{j}\right]=i\epsilon_{ij}^{\phantom{{ij}}k}J_{k}
\end{equation}
I can define a generalized coherent state labelling it with a unitary vector $\hat{n}=\left(\sin\theta\cos\phi,\,\sin\theta\sin\phi,\,\cos\theta\right)$:
\begin{equation}
\left|\hat{n}\right\rangle =\exp\left(i\theta\hat{m}\cdot\vec{J}\right)\left|j,-j\right\rangle =D\left(\hat{n}\right)\left|j,-j\right\rangle
\end{equation}
where $\hat{m}=\left(\sin\phi,\,-\cos\phi,\,0\right)$. Another possible representation use the stereographic projection map $S^{2}\to\mathbb{C}$ to label the coherent state with a complex number. In details if we define $\hat{n}\to\xi=\tan\left(\frac{\theta}{2}\right)e^{i\phi}$ and we call $\tilde{\xi}= \frac{\xi}{\left|\xi\right|} \arctan \left( \xi \right)$ 
\begin{equation}
\left|\hat{n}\right\rangle =\left|\xi\right\rangle =\exp\left(\tilde\xi J^{+}-\bar{\tilde{\xi}}J^{-}\right)\left|j,-j\right\rangle = D\left(\xi\right)\left|j,-j\right\rangle
\end{equation}
These $SU(2)$ coherent states are characterized by the properties:
\begin{itemize}
\item The lowest spin state in the $j$ irrep. is a coherent state with $\hat{n}=\hat{z}$ ($\xi=0$);
\item They belong to the spin $j$ irrep.
\begin{equation}
J^2 \left|\hat{n}\right\rangle = D\left(\hat{n}\right) J^2 \left|j,-j\right\rangle = j\left(j+1\right)\left|\hat{n}\right\rangle \, ;
\end{equation}
\item the displacement operator $D\left(\hat{n}\right)$ is unitary;
\item the action of the displacement operator on the generators rotates them:
\begin{equation}
D^\dagger\left(\hat{n}\right) \vec{J} D \left(\hat{n}\right) = R\left(\hat{z},\hat{n}\right) \vec{J}
\end{equation}
where $R\left(\hat{z},\hat{n}\right)$ is the rotation matrix that transforms $\hat{z}$ into $\hat{n}$;
\item the coherent state $\left|\hat{n}\right\rangle$ is eigenvector of the operator $\hat{n}\cdot\vec{J}$
\begin{equation}
\hat{n}\cdot\vec{J} \left|\hat{n}\right\rangle = -j \left|\hat{n}\right\rangle \, ;
\end{equation}
\item the scalar product of two coherent state reads
\begin{equation}
\left\langle\hat{n}\right|\left.\hat{m}\right\rangle =e^{i\, j\Phi\left(\hat{n},\hat{m}\right)}\left(\frac{1+\hat{n}\cdot\hat{m}}{2}\right)^{j}\, ,
\end{equation} 
where $\Phi$ is the area of the spherical triangle identified by
$\hat{z}$, $\hat{n}$, $\hat{m}$, or
\begin{equation}
\left\langle\xi\right|\left.\eta\right\rangle =\left(\frac{\left(1+\xi\bar{\eta}\right)^{2}}{\left(1+\left|\xi\right|^{2}\right)\left(1+\left|\eta\right|^{2}\right)}\right)^{j}\, ;
\end{equation} 
\item although the coherent states are not orthogonal, they form an over-complete set of states
\begin{equation}
\frac{2j+1}{4\pi}\int_{S^{2}}{\rm d}\cos\theta{\rm d}\phi\left|\hat{n}\right\rangle \left\langle \hat{n}\right|=1 \, ;
\end{equation}
or 
\begin{equation}
\frac{2j+1}{2\pi}\int\frac{{\rm d}\xi\wedge{\rm d}\bar{\xi}}{\left(1+\left|\xi\right|^{2}\right)^{2}}\left|\xi\right\rangle \left\langle \xi\right|=1 \, ;
\end{equation}
\item they minimize the Heisenberg uncertainty inequality. Let's consider three orthogonal unitary vector $\hat{\ell}$, $\hat{m}$, $\hat{n}$, then on the coherent state $\hat{n}$
\begin{equation}
\left\langle \Delta \hat\ell \cdot \vec J\right\rangle \left\langle \Delta \hat m \cdot \vec J\right\rangle = \frac{1}{4}\left\langle \hat{n} \cdot \vec J \right\rangle ^{2} = \frac{j^2}{4}\,.
\end{equation}
\end{itemize}

\section*{Appendix III. Bogolubov transformations}
\noindent
It is remarkable to notice the way ladder operators of the fermionic Hilbert space rotate under the adjoint action of the displacement operator $D(\xi)$. To fully appreciate it, we shall first go back to the bosonic case, discussed in \eqref{btri} and in \eqref{btri2}. Internal transformation on the ladder operators, hence on the Hilbert space, were implemented in the bosonic Hilbert space by the adjoint action of the displacement operator $D(\alpha)$, and amounted to a mere shift of the ladders operators, as specified in \eqref{btri2}. This construction has been shown in Sec. \ref{sh} to naturally emerge while recovering the Schwinger representations of the Lie group $U(1)$ \cite{reviewFujii}. 

For a detailed analysis we address the reader to \cite{BhramaDonaMarciano}, while for the purpose of this study it is enough no notice that the same procedure can be applied to the fermionic Hilbert space, but finding different results. Indeed taking the BCS states, which are Schwinger representations of the SU(2) group, transformations induced by the displacement operator $D(\xi)$ turn out to be now Bogolubov transformations:
\begin{eqnarray}
\tilde{a}=\cos\left(\left|\xi\right|\right)a+\frac{\xi}{\left|\xi\right|}\sin\left(\left|\xi\right|\right)b^{\dagger}\,,
\end{eqnarray}
and
\begin{eqnarray}
\tilde{b}^{\dagger}=\cos\left(\left|\xi\right|\right)b^{\dagger}-\frac{\bar{\xi}}{\left|\xi\right|}\sin\left(\left|\xi\right|\right)b^{\dagger}\,.
\end{eqnarray}
The importance of this transformation, and its relevant physical consequences, will be clarified in   \cite{BhramaDonaMarciano}. For the meantime, we notice that this is crucial to show invariance of the microscopic condensate state under diffeomorphisms.

\section*{Appendix IV. Curvature perturbations}
\noindent
In this action we summarize how the theory of cosmological perturbations works within the standard set-up. We retrace the very same footsteps that led to the definition of the ``curvature perturbation'' variable $\zeta$ (see {\it e.g.} Refs.~\cite{Mukhanov:1990me, zeta1, zeta2}), in order to clarify the origin of the prescription we proposed in Sec. \ref{scp}.

We start reminding that metric perturbations can be cast in the ADM decomposition \cite{ADM} of a generic line element
\begin{eqnarray}
ds^2=N^2 dt^2 -\gamma_{ij} (dx^i + N^i dt) (dx^j + N^j dt)\,,
\end{eqnarray}
in which $N$ denotes the lapse function and $N^i$ the shift vector. A unit time-like vector $n^\mu$ can be defined, which is normal to the hypersurfaces of constant coordinate time $t$ and whose components read
\begin{eqnarray}
n_\mu=(N, 0)\,, \qquad n^\mu=(-\frac{1}{N}, \frac{N^i}{N})\,.
\end{eqnarray}
The extrinsic curvature tensor, which measures how much the hyper-surface is curved in the way it sits in the spacetime manifold, or in other words it measures the failure of a vector tangent to the hyper-surface to remain tangent after parallel transporting it with respect to the Levi-Civita connection on the space-time manifold, reads  
\begin{eqnarray}
K_{ij}&=&-\nabla_{(j} n_{i)}= \\
&=& \frac{1}{2N} \left( -\partial_t \gamma_{ij} + \!\!\!\phantom{a}^{(3)}\nabla_{(i} N_{j)} + \!\!\!\phantom{a}^{(3)}\nabla_{(j} N_{i)}  \right) \nonumber\,,
\end{eqnarray}
in which $\!\!\!\phantom{a}^{(3)}\nabla_i$ refers to the covariant derivatives with respect to the Levi-Civita connection on the spatial hyper-surface. Extrinsic curvature can be decomposed in terms of a symmetric traceless tensor $A_{ij}$, namely $A_{ij}\gamma^{ij}=0$, plus the three-metric tensor itself times a scale quantity $\theta$, namely 
\begin{eqnarray} \label{K}
K_{ij}=-\frac{\theta}{3} \gamma_{ij} +A_{ij}\,.
\end{eqnarray}
The quantity $\theta$ appearing in \eqref{K} represents the volume expansion rate of the spatial hypersurfaces along the integral curves $\gamma(\tau)$ (the proper time $\tau$ is obtained by the definition $d\tau=Ndt$) of $n^\mu$, and is given by $\theta=\nabla_\mu n^\mu$. The number of e-folds of the expansion is therefore expressed, in terms of its dependence on two fixed time-coordinates  of the initial and final hypersurfaces and on the comoving space coordinates $x_i$, as
\begin{eqnarray}
N(t_1,t_2; x_j)=\frac{1}{3} \int_{\gamma(\tau)} \theta d\tau =\frac{1}{3} \int \limits_{t_1}^{t_2} \theta N dt\,.
\end{eqnarray}
The spatial metric $\gamma_{ij}$ can be then decomposed, introducing a local scale factor $a(t, x_i)$, and a unimodular metric $\tilde{\gamma}_{ij}$, namely
\begin{eqnarray}
\gamma_{ij}=a(t, x_i)\, \tilde{\gamma}_{ij}.
\end{eqnarray}
The unimodular metric $\tilde{\gamma}_{ij}$ can be finally expressed in terms of a primordial perturbations tensor, which is a traceless matrix  $h_{ij}$ such that 
\begin{eqnarray}
\gamma_{ij}=(e^h)_{ij}.
\end{eqnarray}
The local scale factor $a(t, x_i)$ can be also decomposed into a global scale factor, which is independent on the position on the space hypersurfaces, and a local deviation $\psi(t,x_i)$, namely
\begin{eqnarray}
a(t, x_i)=a(t) \, e^{\psi(t,x_i)}\,,
\end{eqnarray}
in which such a deviation is assumed to be for our purposes a local (``scalar'') perturbation. In other words, $a(t)$ is chosen in such a way that $\psi(t, x_i)$ vanishes somewhere in the Universe. The gradient expansion method \cite{gep1,gep2,gep3} can be applied in order to expand inhomogeneities into their spatial gradients, and formally multiply them by a fictitious parameter $\epsilon$ regulating the expansion. Following \cite{zeta2}, we may identify the infinitesimal expansion parameter with the ratio between the Hubble radius and a comoving scales of physical size, thus $\epsilon = k/(a \,H)$. Then, on super-horizon scales $A_{ij}=O(\epsilon)$, which allow us to disregard it with respect to quantities referring to a homogenous and isotropic FLRW universe. Since the local expansion recasts as
\begin{eqnarray} \label{htilde}
\theta=\frac{3}{N} \left( \frac{\dot{a}(t)}{a(t)} +\dot{\psi} \right)\equiv 3 \tilde{H}\,,
\end{eqnarray}
having introduced the local Hubble parameter $\tilde{H}\equiv \theta/3$, we find the leading expression for the extrinsic curvature 
\begin{eqnarray}
K^i_{\ j}=-\frac{1}{N} \left( \frac{\dot{a}(t)}{a(t)} +\dot{\psi} \right) \delta^i_j + O(\epsilon)\,,
\end{eqnarray}
and finally, for a conformally flat three-geometry, characterized by $\tilde{\gamma}_{ij}=\delta_{ij}$, the intrinsic curvature on the spatial three-dimensional hypersurfaces is found, which further clarifies the meaning of $\psi$ and its gradients:  
\begin{eqnarray}
\phantom{a}^{(3)} R= -\frac{2}{a^2(t) e^{2\psi}} \delta^{ij} \left( \psi_{,i}  \psi_{,j} + 2  \psi_{,ij} \right)
\,.
\end{eqnarray}
Using the Gau\ss-Codacci equations, the Einstein equations can be recast in terms of $\phantom{a}^{(3)} R$, $K_{\,j}^i$ and $A_{ij}$ (see {\it e.g.} Ref.~\cite{zeta2}), and expanded at linear order in $\epsilon$. 

Within the separate Universe assumption, we can write then the conservation of the energy-momentum tensor at each point, namely $\nabla_\mu T^{\mu \nu}=0$, as 
\begin{eqnarray}
\frac{d\rho(t,x_i)}{dt}= - 3 \tilde{H} (t,x_i) [\rho(t,x_i)+p(t,x_i)]\,,
\end{eqnarray}
which reads the same as in a FLRW universe. Nonetheless, choosing a slicing where $\rho(t,x_i)=\rho(t)$, namely the energy density is uniform, and assuming pressure to be adiabatic, {\it i.e.} to be a unique function of the energy density, namely $p=p(\rho)$, entails the relation 
\begin{eqnarray}
\frac{d\rho(t)}{dt} +  3 \frac{\dot{a}(t)}{a(t)} [\rho(t)+p(t)]= \dot{\psi}(t)\,,
\end{eqnarray}
in which $\psi$ must be independent on the space position. Conservation of the energy-momentum tensor on the FLRW background finally implies conservation of $\psi$, which we denote here as $\zeta$, if the ``adiabatic pressure condition'' is satisfied. Indeed $\zeta$, which determines the intrinsic curvature of constant time spatial hypersurfaces, can be shown to be constant whenever pressure can be expressed as a unique function of the energy density. In particular, this is true in the matter and radiation dominated eras during the expansion of the Universe. Conservation of curvature perturbation then arises when the uniform density slicing coincides at the first order in $\epsilon$ with the comoving and uniform-Hubble slicing, namely the slicing orthogonal to the comoving worldlines. Choosing the comoving worldlines as the threading then fixes the gauge completely to be the so called comoving gauge. This choice of the comoving slice further sets vorticity of the cosmological fluid to zero, consistently with the fact that this latter is not generated during inflation. \\

The curvature perturbation variable $\zeta$ can be evaluated in this framework by linking it to the perturbation of the energy density. Choosing a class of threading in which $N^i=O(\epsilon)$, we may select an initial slice where the energy-density is uniform, and then follow the evolution of the system towards an hyper-surface where its energy-density is not uniform. Concretely, we first estimate the number of e-foldings of expansion along the comoving worldline to which $n^\mu$ is tangent, {\it i.e.}
\begin{eqnarray} \label{deltaN}
N(t_2,t_1;x^i)=\frac{1}{3}\int_{t_1}^{t_2} \theta \, N\,  dt = -\frac{1}{3} \int_{t_1}^{t_2} dt \frac{\dot{\rho}}{\rho+p}\Bigg|_{x^i},\ \
\end{eqnarray}
and then compare two different choices of time slicing, which entail different space-dependence of the energy-density on the hypersurfaces. Thus, first we combine \eqref{htilde} with \eqref{deltaN}, so to obtain 
\begin{eqnarray}  \label{psin}
\psi(t_2,x^i)-\psi(t_1,x^i)=N(t_2,t_1;x^i)-\ln \left( \frac{a(t_2)}{a(t_1)}\right), \ \
\end{eqnarray}
then we deploy the strategy outlined above, and consider two different time-slicing, which coincide at $t=t_1$, and evolve differently up to the hyper-surface at constant time $t=t_2$, where at generic space-positions $x^i$ the perturbation variables differ by
\begin{eqnarray} \label{n1}
\psi_A(t_2,x^i)-\psi_B(t_2,x^i)&=&N_A(t_2,t_1;x^i)-N_B(t_2,t_1;x^i)\nonumber\\
&\equiv&\Delta N_{AB} (t_2,x^i)\,.
\end{eqnarray}
If we choose the A threading to start at a flat slice $t=t_1$, and to end at $t=t_2$ at a uniform density slice, and select the B slicing to be flat at initial time and final time, applying \eqref{n1} we find that $\psi_A(t_2,x^i)=N_A(t_2,t_1; x^i)-N_0(t_2,t_1)$. Applying \eqref{psin} to the adiabatic case $p=p(\rho)$, we finally find 
\begin{eqnarray}
\psi(t_2,x^i)-\psi(t_1,x^i)= -\ln \left[ \frac{a(t_2)}{a(t_1)} \right] -\frac{1}{3} \int^{\rho(t_2,x^i)}_{\rho(t_1,x^i)} \frac{d\rho}{\rho+p}\,. \nonumber
\end{eqnarray}
The latter relation implies the existence of a conserved quantity 
\begin{eqnarray}\noindent 
-\zeta(x^i)=\psi(t,x^i)+ \frac{1}{3} \int_{\rho(t)}^{\rho(t,x^i)} \frac{d\rho}{\rho+p}\,,
\end{eqnarray}
which in the limit of a linear theory, reduces to the expression for the conserved curvature perturbation in the uniform-density, uniform-Hubble, or the comoving slicing, namely 
\begin{eqnarray}
-\zeta(x^i)=\psi(t,x^i)+ \frac{\delta \rho}{3(\rho+p)}\,.
\end{eqnarray}
Notice finally that choosing an arbitrary threading such that $x^i=x^i(t',x^i)$ introduces a generic time dependence in $\zeta$, and in general relaxes the imposition $N^i=O(\epsilon)$.\\

Inhomogeneities of the energy densities can be linked to $\zeta$ also by deploying slightly different arguments, closely related to the $\delta N$ formalism. For the purpose of this paper we will expose this latter too, since it adapts to our arguments on the generalization of cosmological perturbations to the quantum realm. Again we consider a change from a uniform density slicing to any other generic one. And again we emphasize that on super horizon scales the threading is uniquely defined, and a change of slicing only amounts to a shift in the coordinate time. Thus, at any given position this change entails a time change $\delta(t,x^i)$ such that $t'=t+\delta t(t,x^i)$. Correspondingly, if we keep fixed points in the background manifold and investigate the change in the
mapping to the perturbed manifold, we find for the expression of the local scale factor 
\begin{eqnarray}
a(t',x^i)=a(t,x)-\dot{a}(t)\delta t\,.
\end{eqnarray}
We then separate the local scale factors in a background part and a perturbative part, using $a(t,x^i)=a(t) e^{\psi(t,x^i)}$, and recover for small perturbations that  
\begin{eqnarray} \label{ps1}
\psi=\zeta-H \delta t\,.
\end{eqnarray}
In a similar way we may recover for the  perturbations on the energy density the relation
\begin{eqnarray} \label{ps2}
\delta \rho_\psi(t,x^i)=-\dot{\rho}(t) \,\delta(t,x^i)\,,
\end{eqnarray}
which has been evaluated on a uniform density slicing, characterize by $\delta \rho=0$, and on an arbitrary slicing $\delta_\psi$. Combining \eqref{ps1} with \eqref{ps2} finally give us
\begin{eqnarray} \label{ps23}
\zeta=\psi- \frac{\delta \rho_\psi(t,x^i)}{\rho+p}\,.
\end{eqnarray}

\section*{Appendix V. Scale invariant power spectrum}

\noindent 
In this section we summarize how scale invariant power spectra can be originated by curvature perturbations in the scenario of cosmological inflation, and focus on the paradigmatic case represented by the so-called slow-roll approximation. 

The main inspiring idea is that the field is slowly rolling toward the bottom of its potential well $V(\phi)$, so slowly that its kinetic energy $K_\phi=\dot{\phi}^2/2$ is negligible with respect to its potential energy. Therefore the value of $\phi$ is almost constant, which localizes it at a certain $\phi_0$ on the potential well. We then have
\begin{eqnarray}
\rho\simeq V(\phi)\, \qquad {\rm and} \qquad 3H\dot{\phi}\simeq V'\,.
\end{eqnarray}
On the other side, this can happen only if the potential is flat ``enough'' to allow for this approximation, which is a requirement on its derivatives in $\phi$. One is finally led to introduce the slow-roll parameters in cosmological inflation, which we denote here as $\epsilon_{\rm s.r.}$ and $\eta_{\rm s.r.}$ and read 
\begin{eqnarray}
\epsilon_{\rm s.r.}\equiv \frac{m^2_{\rm Pl}}{2}\, \left( \frac{V'}{V} \right)^2, \quad {\rm and} \quad \eta_{\rm s.r.}\equiv m^2_{\rm Pl}\, \frac{V''}{V}\,. 
\end{eqnarray}
To be fully specific, we can limit our focus to the relevant case of quadratic potentials, namely to stochastic inflation \cite{Linde}. After the suitable redefinition $\chi=a(\eta) \phi$, the equation of motion for the background field $\chi$ can be cast in conformal coordinates $\{\eta, x^i\}$, which are such that $ds^2=a(\eta)^2(d\eta^2-d\vec{x}^2)$, as
\begin{eqnarray} \label{sfr}
\chi''-\nabla^2 \chi + \left( a^2 m^2 \!-\!\frac{a''}{a} \right) \chi =0\,.
\end{eqnarray}
Exactly the same expression as in \eqref{sfr} holds for the variation $\delta \chi= a(\eta) \delta \phi$. This is the well celebrated equation for the cosmological perturbations, which have a tachyonic mass. Quantization of $\delta \chi$ perturbations now proceed in a similar way as for the $\delta \phi$, showing exactly the same expansion as in \eqref{furdf}, and again involving commutation relations. We may focus now on the perturbation $\delta\chi$, whose equation of motion for the Fourier space-modes reads
\begin{eqnarray}
\delta \chi_{\vec{k}}(\eta)''\!+\!(a H)^2 \!\left[  \left( \frac{m}{H} \right)^2 \!+\! \left( \frac{k}{a H} \right)^2 \!\!-\!\frac{H'}{H}\!-\!2 \right] \!\delta\chi_{\vec k}(\eta) = 0,\nonumber
\end{eqnarray}
in which we also have used $a''/a^3=H'+2H^2$. A solution to the latter equation, which is found for the de Sitter background and fulfills plane-waves initial conditions, namely is matched to the Bunch-Davies vacuum \cite{Bunch:1978yq}, involves Hankel functions and reads
\begin{eqnarray}
\delta \chi_{\vec{k}}(\eta)=\sqrt{ \frac{-\eta \pi}{2}} \, e^{\imath \frac{\pi}{4} (2 \nu +1) }\, H_\nu^{(1)} (-k \eta)\,.
\end{eqnarray}
The specific details of the model are now contained in  
\begin{eqnarray}
\nu=\sqrt{\frac{9}{4}-\left( \frac{m}{H}\right)^2}\,.
\end{eqnarray}
Recalling that in de Sitter $\eta=-(a H)^{-1}$, well after crossing the Hubble horizon, {\it i.e.} for super-horizon modes fulfilling $|k \eta|\!<\!\!<\!1$ 
the solution approaches 
\begin{eqnarray}
\delta \chi_{\vec{k}}(\eta)= \frac{e^{\imath \frac{\pi}{4} (2\nu -1) }}{\sqrt{2 k }}\, \frac{\Gamma(\nu)}{\sqrt{\pi}}\, \left( \frac{-k \eta}{2}\right)^{\frac{1}{2}-\nu}\,.
\end{eqnarray}
The power spectrum can be computed from the correlation function of the perturbation variables $\delta\chi$ using the formula
\begin{eqnarray}
&&\!\!\!\!\!\! \langle 0| \hat{\delta \chi}(\eta, \vec{x}) \hat{\delta \chi}(\eta, \vec{y}) |0 \rangle =\\
&&\!\!\!\!=\frac{1}{2\pi^2} \int_0^\infty\!\!\!\! k^3 |\delta \chi_{\vec{k}}(\eta)|^2\, \frac{\sin kL}{k L}\, \frac{dk}{k}
\equiv \int_0^\infty \!\!\!\! k^3 \mathcal{P}_{\delta\chi}\, \frac{\sin kL}{k L}\, \frac{dk}{k}, \nonumber
\end{eqnarray}
in which we have introduced the coordinate space distance $L=|\vec{x}-\vec{y}|$ and defined the power spectrum  $\mathcal{P}_{\delta\chi}= (k^3/2\pi^2)|\delta \chi_{\vec{k}}|^2$. Experimentally $L$ individuates a pivotal scale, which we will call later $k_0$.

Notice that scale-invariance is attained whenever $\nu=3/2$. The power spectrum can be then recast in terms of $\mathcal{P}_{\delta\phi}$, by considering that $\delta \phi = \delta \chi/a$. The change of variables allows to find for $\nu=3/2$
\begin{eqnarray}
\mathcal{P}_{\delta\phi}=\left(\frac{H}{2\pi} \right)^2\,.
\end{eqnarray}
It is straightforward to check that scale-invariance of the power-spectrum implies $\delta \chi_{\vec{k}}\simeq k^{-\frac{3}{2}}$. Furthermore, for inflation $H\simeq H_k$, and the Hubble parameter can finally be evaluated at the horizon-exit, looking for modes $a H_k=k$, to be $H\simeq 10^{-5}$ in Planckian units. This estimate may already give a realistic value for the power spectrum of scalar perturbations. Slight deviations from scale invariance are then parametrized for light field $m\leq 3H/2$ by 
\begin{eqnarray}
\nu \simeq \frac{3}{2} -\frac{m^2}{3 H^2}\,, 
\end{eqnarray}
which entails 
\begin{eqnarray}
\mathcal{P}_{\delta\phi}=\left(\frac{H}{2\pi} \right)^2\, \left(\frac{k}{2 a H} \right)^{\frac{2}{3} (\frac{m}{H})^2}\,.
\end{eqnarray}
Whenever a generic dependence in $k$ is present, we can define a corresponding spectral index for the power spectrum by
\begin{eqnarray}
n-1\equiv \frac{d \ln \mathcal{P}}{d \ln k}\,.
\end{eqnarray}
The power spectrum of scalar perturbations is finally recovered by noting that for slow-roll inflation $\rho\simeq V(\phi)$ and $3 H \dot{\phi}\simeq - V'(\phi)$ hold, and thus 
\begin{eqnarray}
\zeta=\frac{1}{3}\frac{V'}{\dot{\phi}^2}\delta \phi = \frac{1}{m_{\rm Pl}^2}\frac{V}{V'} \delta \phi\,.
\end{eqnarray}
For almost scale invariant massive fields, {\it i.e.} for $\nu=3/2$, one then finds 
\begin{eqnarray}
\mathcal{P}_{\zeta}= \left(  \frac{1}{m_{\rm Pl}^2}\frac{V}{V'}  \right)^2 \!\!\left( \frac{H_k}{2\pi}\right)^2\!\!\simeq \frac{1}{24 m_{\rm Pl}^4 \pi^2 }\,\frac{V}{\epsilon_{\rm s.r.}}\,,
\end{eqnarray}
which involves the slow roll parameter $\epsilon_{\rm s.r.}$ introduced above, and must be evaluated at the horizon exit. The experimental value of the scalar power spectrum to be used in constraining parameters in $\mathcal{P}_{\zeta}$ is expressed \cite{WMAP, Planck} by $\mathcal{P}_\zeta(k_0)=(2.445\pm0.096)\times 10^{-9}$, in which the pivotal scale chosen corresponds to $k_0=0.002{\rm Mpc}^{-1}$. The power spectrum is finally expressed as 
\begin{eqnarray}
\mathcal{P}_\zeta(k)=\mathcal{P}_\zeta(k_0)\left( \frac{k}{k_0}\right)^{n}\,,
\end{eqnarray}
if, assuming independence on $k$, the spectral index is fitted with experimental data to be
\begin{eqnarray}
n=0.960\pm0.013\,.
\end{eqnarray}

\section*{Acknowledgment}
\noindent
We acknowledge useful criticisms by M.~Sasaki during LeCosPa II symposium, and enlightening comments by N.~Bartolo, S.~Brahma, Y.~Cai, F.~Cianfrani, A.~Hamma, L.~Hui, S.~Matarrese, F.~Piazza, C.~Rovelli and E.~Wilson-Ewing over the early development of these ideas. P.D. and A.M. wishes also to thank colleagues working in condensed matter within the department of physics of Fudan University for inspiring discussions, as well as Dr.~A.P.~Wine for very much inspiration during the early drafting of this note.

\end{document}